\documentclass[11pt]{article}
\pdfoutput=1
\usepackage{jcapmod}
\usepackage{shorthand}
\usepackage[scaled]{helvet}
\usepackage[toc,page]{appendix}
\usepackage{mathtools}
\usepackage{booktabs}
\usepackage[english]{babel}
\usepackage{amsmath,amssymb,amsbsy,amstext, amsthm, simplewick, amsfonts}
\usepackage{hyperref}
\usepackage{graphicx}
\usepackage[small]{caption}
\usepackage{siunitx}
\usepackage{upgreek}
\usepackage{framed}
\usepackage{wrapfig}
\usepackage{multirow}
\usepackage{tabularx}
\usepackage{bbm}
\usepackage[svgnames,dvipsnames,x11names]{xcolor}
\usepackage[utf8x]{inputenc}
\usepackage{selinput}
\usepackage{bm}
\usepackage{float}
\usepackage{geometry}
\usepackage{yfonts}
\usepackage{caption}
\usepackage{subcaption}
\usepackage{sidecap}
\usepackage{longtable}
\usepackage{anyfontsize}
\usepackage{dsfont}
\usepackage{tikz}
\usepackage{relsize}
\usepackage{tcolorbox}
\usepackage{shuffle}
\usetikzlibrary{matrix}
\usepackage{bigstrut}
\usepackage{fnpct}
\usetikzlibrary{arrows.meta}
\usetikzlibrary{decorations.pathreplacing,calligraphy}
\usepackage{tensor}
\usepackage{mleftright}

\usetikzlibrary{decorations.markings, decorations.pathmorphing, decorations.pathreplacing, decorations.shapes}

\usepackage{xcolor}
\usepackage{xparse}

\usepackage{fontawesome5}

\makeatletter
\newcommand{\github}[1]{%
   \href{#1}{\faGithub}%
}
\makeatother

\NewDocumentCommand{\colornucleus}{omme{_^}}{%
  \begingroup\colorlet{currcolor}{.}%
  \IfValueTF{#1}
   {\textcolor[#1]{#2}}
   {\textcolor{#2}}
    {%
     #3
     \IfValueT{#4}{_{\textcolor{currcolor}{#4}}}
     \IfValueT{#5}{^{\textcolor{currcolor}{#5}}}
    }%
  \endgroup
}

\usepackage{array}
\newcolumntype{L}[1]{>{\raggedright\let\newline\\\arraybackslash\hspace{0pt}}m{#1}}
\newcolumntype{C}[1]{>{\centering\let\newline\\\arraybackslash\hspace{0pt}}m{#1}}
\newcolumntype{R}[1]{>{\raggedleft\let\newline\\\arraybackslash\hspace{0pt}}m{#1}}

\usepackage[framemethod=default]{mdframed}
\newmdenv[skipabove=7pt,
skipbelow=7pt,
rightline=true,
leftline=true,
topline=true,
bottomline=true,
backgroundcolor=gray!10,
linecolor=black,
innerleftmargin=5pt,
innerrightmargin=5pt,
innertopmargin=5pt,
innerbottommargin=5pt,
leftmargin=0cm,
rightmargin=0cm,
linewidth=1pt]{eBox}

\definecolor{Red}{RGB}{214, 39, 40}
\definecolor{Blue}{RGB} {31, 119, 180}
\definecolor{Orange}{RGB}{255, 153, 51}
\definecolor{Purple}{RGB}{178, 102, 255}
\definecolor{Green}{RGB}{44, 160, 44}
\definecolor{regal}{RGB}{90,0,120}
\definecolor{darkblue}{rgb}{0.15,0.35,0.55}
\definecolor{reddish}{rgb}{0.65, 0.2, 0.2}
\definecolor{darkgreen}{RGB}{50,150,0}
\definecolor{greyish}{rgb}{.90,.90,.90}
\definecolor{greyish2}{rgb}{.96,.96,.96}
\definecolor{greyish3}{rgb}{.37,.37,.37}
\definecolor{darkblue2}{rgb}{0.3,0.4,0.9}
\definecolor{Blue3}{RGB}{31, 119, 180}

\definecolor{blue3}{RGB}{31, 119, 180}
\definecolor{red3}{RGB}{	214, 39, 40}
\definecolor{orange3}{RGB}{255, 127, 14}
\definecolor{green3}{RGB}{44, 160, 44}
\definecolor{repBlue}{RGB}{31, 119, 180}
\definecolor{repRed}{RGB}{	214, 39, 40}
\definecolor{repGreen}{RGB}{44, 160, 44}

\definecolor{vio}{RGB}{19, 130, 164}
\definecolor{vioo}{RGB}{89, 2, 155}
\newcommand{\Comment}[1]{{}}
\usepackage[linktocpage=true]{hyperref}
\hypersetup{
colorlinks=true,
citecolor=darkblue,
linkcolor=reddish,
urlcolor=darkblue,
pdfauthor={},
pdftitle={},
pdfsubject={}
}

\usepackage{colortbl}
\definecolor{lightgreen}{cmyk}{0.2, 0, 0.2, 0.2}
\definecolor{lightgray2}{cmyk}{0.1,0.1,0,0.1}
\definecolor{Red2}{RGB}{214, 39, 40}
\definecolor{Blue2}{RGB} {31, 119, 180}
\definecolor{Orange2}{RGB}{255, 127, 14}
\definecolor{Green2}{RGB}{44, 160, 44}

\makeatletter
\newlength{\apb@width}
\newcommand{\autoparbox}[2][c]{\settowidth{\apb@width}{#2}\parbox[#1]{\apb@width}{#2}}

\makeatother


\def\beq{\begin{equation}}
\def\eeq{\end{equation}}
\def\be{\begin{equation}}
\def\ee{\end{equation}}

\allowdisplaybreaks[1]
\newcommand\sqmatrix[2][c]{%
  \fixTABwidth{T}%
  \setbox0=\hbox{$\tabbedCenterstack{#2}$}%
  \setstackgap{L}{\dimexpr\maxTAB@width+\tabbed@gap}%
  \tabbedCenterstack[#1]{#2}%
}

\usetikzlibrary{shapes.misc}

\tikzset{cross/.style={cross out, draw=black, minimum size=2*(#1-\pgflinewidth), inner sep=0pt, outer sep=0pt},
cross/.default={1pt}}

\usetikzlibrary{calc}
\def\be{\begin{eqnarray}}
\def\ee{\end{eqnarray}}
\def\no{\nonumber}
\def\vs{\nonumber\\ }
\def\VEV#1{\left\langle #1 \right\rangle}
\renewcommand{\refeq}[1]{Eq.~(\ref{eq:#1})}
\newcommand{\refeqs}[2]{Eqs.~(\ref{eq:#1})--(\ref{eq:#2})}
\newcommand{\reffig}[1]{Fig.~\ref{fig:#1}}
\newcommand{\refsec}[1]{Sec.~\ref{sec:#1}}
\newcommand{\refapp}[1]{App.~\ref{app:#1}}
\newcommand{\reftab}[1]{Table~\ref{tab:#1}}

\def\bx{{\bf x}}
\def\by{{\bf y}}
\def\TL{{\rm TL}}
\def\dphi{\delta\phi}
\def\dpi{\delta\pi}
\def\calQ{\mathcal{Q}}
\def\calO{\mathcal{O}}
\def\calN{\mathcal{N}}

\begin{document}
\begin{titlepage}
\setcounter{page}{1} \baselineskip=15.5pt 
\thispagestyle{empty}

\begin{center}
{\fontsize{30}{30} \bf Quantum Popcorn}\vskip 4pt
{\fontsize{16}{16} \bf The shape of a phase transition bubble}
\end{center}

\vskip 30pt
\begin{center}
\noindent
{\fontsize{12}{18}\selectfont 
Kfir Blum$^1$, Ipak Fadakar$^2$, Omri Rosner$^1$, and Guilherme L.~Pimentel$^2$}
\end{center}

\begin{center}
\vskip 4pt
\textit{$^1$Weizmann Institute of Science, Rehovot 7610001, Israel}\\ \vskip 4pt
\textit{$^2$Scuola Normale Superiore and INFN, Piazza dei Cavalieri 7, 56126, Pisa, Italy}
\end{center}

\vspace{0.4cm}
\begin{center}{\bf Abstract}\end{center}
A quantum tunneling event in field theory is usually depicted as the materialization of a spherical bubble of true vacuum. The bubble appears at rest, and its profile is the time-symmetric slice of Coleman's bounce, the Euclidean solution that controls the decay rate. However, the bubble is spherical only in the leading semiclassical approximation: at the next order the configuration is a Gaussian ensemble of fluctuations around the bounce. We derive the wave functional of the field on the nucleation slice from the half-bounce path integral, and use it to compute the root-mean-square quadrupole moment of a bubble at nucleation. The relative deformation of the bubble wall radius $R$ is $\delta R/R\sim(\hbar/S_b)^{1/2}$, with $S_b$ the bounce action. Thus $\delta R/R\sim0.1$ for a typical cosmological phase transition with $S_b/\hbar\sim100$.
\noindent

\end{titlepage}

\newpage
\tableofcontents
\section{Introduction}\label{sec:int}
First-order cosmological phase transitions in field theory are thought to take place in many extensions of the Standard Model, including theories of grand unification~\cite{Guth:1981uk,Guth:1982pn}, solutions of the strong CP problem~\cite{DelleRose:2019pgi,VonHarling:2019rgb}, extra dimensions~\cite{Creminelli:2001th,Randall:2006py,Konstandin:2011dr}, and supersymmetry~\cite{Carena:1996wj,Apreda:2001us}. They can be quantum~\cite{Coleman:1977py} or thermal~\cite{Linde:1980tt,Linde:1981zj,Affleck:1980ac}: here we focus on the former kind. A quantum first-order phase transition typically involves a rare tunneling event wherein a microscopic bubble of true vacuum materializes in a sea of false vacuum~\cite{Coleman:1977py,Callan:1977pt}. The bubble nucleation event by itself, being both exceedingly rare and microscopic, makes little immediate impact on the state of the Universe at large. However, once nucleated, the little bubble proceeds to grow at the speed of light (or near it), converting macroscopic swaths of false vacuum to true until it comes to rule the world. At this later stage of the evolution bubble collisions become probable, and are considered a possible source of a stochastic background of gravitational waves~\cite{10.1093/mnras/218.4.629,Kamionkowski:1993fg,Weir:2017wfa}. 

Because the most phenomenologically consequential dynamics in a phase transition likely takes place during the final stages of bubble expansion and percolation, relatively little attention has been directed to the details of the nucleation event itself. The tunneling rate can be calculated with semiclassical methods, following the work of Coleman and Callan \& Coleman~\cite{Coleman:1977py,Callan:1977pt}, and it is thought that the minimum-action (and thus maximum-probability) tunneling field configuration is spherical~\cite{Coleman:1977th,Blum:2016ipp}\footnote{At least in theories involving only scalar fields; for related discussion see~\cite{Avraham:2026mgk}.}, so it cannot, for example, produce gravitational waves. But other than that, the physical picture of the tunneling event has remained rather vague.

In this paper we put the microscope, so to speak, on the nascent bubble. Specifically, we ask what the bubble looks like around its materialization time, should one have been able to go back in time and take their microscope to a nucleation site. As often happens with a new thing under the microscope, the view turns out to be fascinating: indeed, the usually considered spherical bubble is only the leading term of a semiclassical expansion. At the next order the field is not a single configuration but a Gaussian ensemble around the ``bounce,''\footnote{We will review the bounce solution shortly. It is the leading saddle solution that controls the tunneling rate, and is a Euclidean-time configuration. A slice of this configuration serves as an initial condition in field space for bubble formation.} and this implies that no nucleating bubble is exactly spherical. To illustrate this point, we compute the root-mean-square (RMS) quadrupole moment of the energy density of a bubble at the moment of nucleation. This analysis also allows us to make contact with the work of Ref.~\cite{Blum:2024hcs}, which considered gravitational wave emission by quantum fluctuations of an expanding bubble. 

\begin{figure}[h!]
\centering
\includegraphics[width=0.7\textwidth]{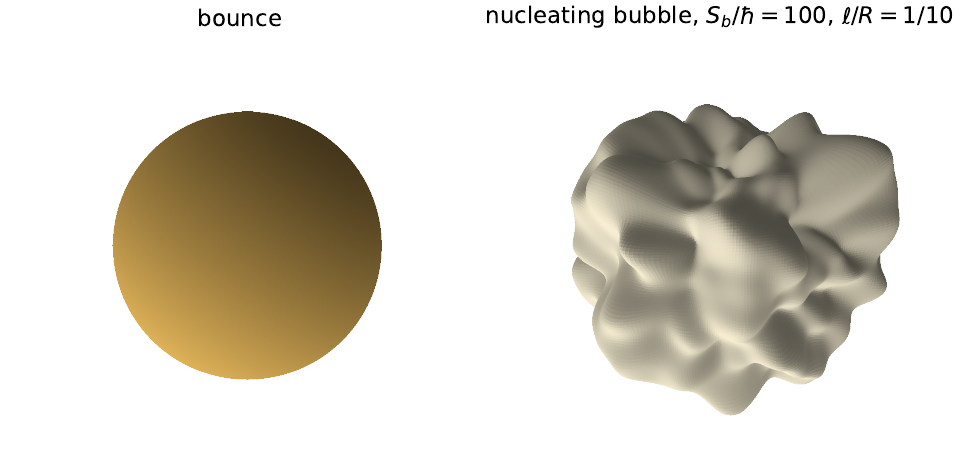}
\caption{The spherical bounce (left) and one bubble drawn from the Gaussian ensemble of the field at nucleation (right), for a thin wall with bounce action $S_b/\hbar=100$. The position of the wall is smoothed on the angular scale $\ell/R$, with $\ell=R/10$, and its root-mean-square displacement is $\delta R/R=0.17$. The construction is explained in \refapp{popcorn}.}
\label{fig:popcorn}
\end{figure}

Up to an order-one model-dependent coefficient, we find that the RMS relative deformation of a bubble wall of radius $R$ is $\delta R/R\sim(\hbar/S_b)^{1/2}$, with $S_b$ the bounce action. For a cosmological phase transition $S_b/\hbar\sim100$, so $\delta R/R\sim10\%$. 
\reffig{popcorn} visualizes this result. It shows the spherical bounce next to one bubble drawn from the Gaussian ensemble at nucleation, for a thin wall with $S_b/\hbar=100$. (The figure shows the wall smoothed on the angular scale $\ell/R$, with $\ell=R/10$; see \refapp{popcorn}). 

The state of the field at nucleation is obtained by cutting the Euclidean bounce at its symmetric slice $\tau=0$. This gives a Gaussian wave functional whose width is set by the Dirichlet-to-Neumann map of the half-bounce. We show in \refsec{state} that the equal-time two-point function of the field in this state is the Euclidean Green's function of the bounce fluctuation operator with both points on the slice. A spectral decomposition of the Green's function then suggests a natural probabilistic interpretation of Callan \& Coleman's fluctuation integral around the bounce~\cite{Callan:1977pt}. 

In \refsec{quad} we expand the energy density to first order in the fluctuations, derive the quadrupole moment at nucleation as the $L=2$ angular moment of the field fluctuation weighted by the wall profile, and obtain its variance as a sum of integrals of Green's function spherical harmonic coefficients weighted by the wall profile.  
We identify the contribution of the wall-displacement mode, and obtain it in closed form for a generic bubble wall profile. 

In \refsec{model} we work out concrete examples. For a thin wall, the wall-displacement mode reproduces the Hartle--Hawking state of the world-volume theory used in Ref.~\cite{Blum:2024hcs}. For wall thickness $\ell_w$ at nucleation radius $R$, in a false vacuum of mass $m$ satisfying $mR\gg1$, internal excitations of the wall are suppressed by $(\ell_w/R)^3$. For a Fubini-like bounce with a ``light'' false vacuum $mR\ll1$, with a bubble profile that extends as $\rho^{-2}$ out to $\rho\sim1/m$, the wall excitation modes are instead enhanced by $\ln(1/mR)$. Both limits are checked against numerical results for a family of quartic potentials from thick to thin walls and for a family of sextic potentials with a light false vacuum. 

In \refsec{disc} we summarize our results. 
Some details of hyperspherical harmonics are deferred to \refapp{hsh}. In \refapp{popcorn} we give a recipe for making quantum popcorn.


\section{The state of the field at nucleation}\label{sec:state}
In this section we work out the statistics of the field configuration near the bounce solution, pointing out that the typical realization will break spherical symmetry.
\subsection{Bounce and fluctuation operator}\label{ss:bounce}
Consider a real scalar field with Euclidean action
\be\label{eq:S} S[\phi]&=&\int d^4x\left[\frac{1}{2}\left(\partial_\mu\phi\right)^2+V(\phi)\right],\ee
with a potential $V$ that has a false vacuum at $\phi=\phi_{\rm FV}$, $V(\phi_{\rm FV})=0$, $V''(\phi_{\rm FV})=m^2$, and a true vacuum with $V<0$. Euclidean coordinates are $x=(\tau,\bx)$, $r=\sqrt{\tau^2+\bx^2}$, $\rho=|\bx|$ and $\hat n=\bx/\rho$. The bounce $\phi_b(r)$ is the $O(4)$-symmetric solution of
\be\label{eq:bounce}\phi_b''+\frac{3}{r}\phi_b'&=&V'(\phi_b),\;\;\;\;\phi_b'(0)=0,\;\;\;\;\phi_b(\infty)=\phi_{\rm FV},\ee
with action $S_b=S[\phi_b]$. It is the determining factor of the probability for the appearance of a bubble of true vacuum embedded in the false-vacuum sea, with bubble nucleation rate per unit volume~\cite{Coleman:1977py,Callan:1977pt}
\be\label{eq:rate}\frac{\Gamma}{V_3}&=&\left(\frac{S_b}{2\pi\hbar}\right)^2\left|\frac{{\rm Det}'\,\calO}{{\rm Det}\,\calO_0}\right|^{-\frac{1}{2}}e^{-S_b/\hbar}\left[1+O(\hbar)\right].\ee
Here
\be\label{eq:O}\calO&=&-\partial^2+V''(\phi_b(r)),\;\;\;\;\calO_0\,=\,-\partial^2+m^2,\ee
is the fluctuation operator, the prime on the determinant removes the four translation zero modes $\partial_\mu\phi_b$, and the absolute value refers to the single negative mode. Expanding $\phi=\phi_b+\dphi$, the one-loop factor is the Gaussian integral
\be\label{eq:gauss}\left|{\rm Det}'\,\calO\right|^{-\frac{1}{2}}&\propto&\int D\dphi\,\exp\left[-\frac{1}{2\hbar}\int d^4x\,\dphi\,\calO\,\dphi\right],\ee
where the $'$ on ${\rm Det}'$ means that the path integral is restricted to the modes with nonzero eigenvalue and with the negative-mode integral rotated to the upward section of the imaginary axis. 

\refeq{gauss} has a natural probabilistic interpretation: it suggests that, conditional on a bubble having nucleated centered on some given point in spacetime, the probability distribution for the bubble to be decorated by a fluctuation $\delta\phi$ is Gaussian, organized by the eigenmodes and eigenvalues of $\calO$. This interpretation is further supported by noting that in \refeq{rate} the spatial volume $V_3$ and the time span $T$ (implicitly defined by $\Gamma$ being probability per unit time) are generated as zero eigenmodes of the same Gaussian integral. The simple Gaussian interpretation is slightly naive, as can be guessed from the negative mode of $\calO$. Nevertheless, we will show in the next sections that with a minor refinement it is essentially correct.

\subsection{The wave functional on the nucleation slice}\label{ssec:wf}
On the slice $\tau=0$, $\partial_\tau\phi_b=(\tau/r)\phi_b'=0$ and the field configuration at leading semiclassical order is $\phi_b(\rho)$, the nucleated bubble at rest. The bubble's subsequent real-time evolution after nucleation is the analytic continuation of the bounce, $\phi_b(\sqrt{\rho^2-t^2})$~\cite{Coleman:1977py}. We define the state of the field at nucleation by the Euclidean path integral over the half space $\tau\leq0$ with the false vacuum at $\tau\to-\infty$ and prescribed field $\phi_0(\bx)$ on the slice,
%
\be\label{eq:Psidef}\Psi[\phi_0]&=&\int\limits_{\phi(-\infty,\bx)=\phi_{\rm FV}}^{\phi(0,\bx)=\phi_0(\bx)}D\phi\;\exp\left(-\frac{S_-[\phi]}{\hbar}\right),\;\;\;\;S_-[\phi]\,=\,\int_{\tau<0}d^4x\left[\frac{1}{2}\left(\partial_\mu\phi\right)^2+V(\phi)\right].\ee
%
Up to normalization, $\Psi[\phi_0]$ is the Schr\"odinger wave functional of the (quasi-stationary) false-vacuum state~\cite{Sasaki:1993gy,Tanaka:1994qa}. For configurations $\phi_0=\phi_b(\rho)+\chi(\bx)$ close to the nucleated bubble, the path integral is dominated by the half-bounce $\phi_b(r)$, $\tau\leq0$, and we expand $\phi=\phi_b+\dphi$ with $\dphi(0,\bx)=\chi(\bx)$. The term linear in $\dphi$ vanishes, $S_1[\dphi]=\int_{\tau<0}d^4x\,[\partial\phi_b\cdot\partial\dphi+V'(\phi_b)\dphi]=\int d^3x\,\dphi\,\partial_\tau\phi_b|_{\tau=0}=0$, because $\partial_\tau\phi_b=0$ on the slice. The quadratic term is
\be\label{eq:S2}S_2[\dphi]&=&\frac{1}{2}\int_{\tau<0}d^4x\left[\left(\partial_\mu\dphi\right)^2+V''(\phi_b)\,\dphi^2\right].\ee
We will compute the statistics of field fluctuations conditional on a bubble having nucleated at $\bx=0$ at $t=0$.

Let $\eta$ be the solution of the fluctuation equation on the half space with prescribed boundary data\footnote{For related analyses, see~\cite{Banks:1973ps,Banks:1973uca,Gervais:1977nv,Bitar:1978vx,Rubakov:1984pa,Vachaspati:1991tq,Sasaki:1993gy,Tanaka:1994qa}.},
\be\label{eq:eta}\calO\,\eta\,=\,0\;\;(\tau<0),\;\;\;\;\eta(0,\bx)\,=\,\chi(\bx),\;\;\;\;\eta\to0\;\;(\tau\to-\infty).\ee
A solution of \refeq{eta} exists for suitable $\chi$ subject to an admissibility condition that we discuss in the next section; given such $\chi$ we can write $\dphi=\eta+\xi$. Here $\dphi$, $\eta$ and $\xi$ are fields on the half space $\tau\leq0$, while $\chi$ is a field on the nucleation slice $\tau=0$. The field $\xi$ carries no slice data and satisfies the boundary conditions
\be\label{eq:xibc}\xi(\tau\to0,\bx)\,=\,\xi(\tau\to-\infty,\bx)\,=\,\xi(\tau,|\bx|\to\infty)\,=\,0.\ee
The cross term between $\eta$ and $\xi$ in $S_2$ vanishes: writing $\partial_\mu\eta\,\partial_\mu\xi=\partial_\mu(\xi\,\partial_\mu\eta)-\xi\,\partial^2\eta$ and applying the divergence theorem on the half space, whose boundary is the slice $\tau=0$ with outward normal $+\hat\tau$,
%
\be\label{eq:cross}\int_{\tau<0}d^4x\left[\partial_\mu\eta\,\partial_\mu\xi+V''(\phi_b)\,\eta\,\xi\right]&=&\int d^3x\;\xi(0,\bx)\,\partial_\tau\eta(0^-,\bx)\;+\;\int_{\tau<0}d^4x\;\xi\,\calO\eta\;=\;0,\ee
%
the first term by \refeq{xibc} and the second by \refeq{eta}. Hence $S_2[\dphi]=S_2[\eta]+S_2[\xi]$. The $\xi$ integral is independent of $\chi$ and can be absorbed in the normalization of the wave functional. The $\eta$ integral follows from integration by parts, using $\eta(0,\bx)=\chi(\bx)$ and $\calO\eta=0$:
\be\label{eq:S2eta}S_2[\eta]&=&\frac{1}{2}\int_{\tau<0}d^4x\left[\left(\partial\eta\right)^2+V''(\phi_b)\,\eta^2\right]\;=\;\frac{1}{2}\int d^3x\,\chi(\bx)\,\left(M\chi\right)(\bx),\vs
\left(M\chi\right)(\bx)&\equiv&\partial_\tau\eta(0^-,\bx).\ee

$M$ is the Dirichlet-to-Neumann map of the half-bounce~\cite{Taylor:dton,Behrndt:2015dtn}. It is a real, symmetric operator on functions on the slice, because \refeq{S2eta} reads $\int d^3x\,\chi_1 M\chi_2=\int_{\tau<0}d^4x\,[\partial\eta_1\cdot\partial\eta_2+V''(\phi_b)\,\eta_1\eta_2]$. The wave functional of the tunneling state at nucleation is therefore the Gaussian~\cite{Sasaki:1993gy,Tanaka:1994qa} 
\be\label{eq:PsiT}\Psi_T[\chi]&=&\calN\,\exp\left[-\frac{S_b}{2\hbar}-\frac{1}{2\hbar}\int d^3x\,\chi\,M\chi\right],\ee
with $\calN$ a normalization constant. 
The momentum conjugate to the slice field $\chi(\bx)$ is the slice operator $\dpi(\bx)\equiv\partial_t\dphi(t,\bx)|_{t=0}=-i\hbar\,\delta/\delta\chi(\bx)$, and on the real Gaussian \refeq{PsiT} it acts as $\dpi\,\Psi_T=i\,(M\chi)\,\Psi_T$. The equal-time covariances on the slice are therefore
%
\be\label{eq:cov}\VEV{\chi(\bx)\chi(\by)}&=&\frac{\hbar}{2}\,M^{-1}(\bx,\by),\;\;\;\;\;\VEV{\dpi(\bx)\dpi(\by)}\,=\,\frac{\hbar}{2}\,M(\bx,\by),\;\;\;\;\;\VEV{\left\{\chi(\bx),\dpi(\by)\right\}}\,=\,0,\ee
%
with $\{\,,\}$ the anticommutator. 
The bubble is ``at rest'' on the slice in the sense that $\VEV{\dpi}=0$ and there is no position--momentum correlation. 
On the slice data on which $M$ is positive, the momentum has the zero-point spread $\hbar M/2$. 

\subsection{Equal-time correlators, and probabilistic interpretation}\label{ssec:glue}
Let $F[\chi]$ be a functional of the slice field. The full-space Gaussian integral \refeq{gauss} with the insertion $F[\dphi(0,\cdot)]$ can be evaluated by integrating first over the fluctuations in $\tau<0$ and in $\tau>0$ at fixed slice value $\chi$, and then over $\chi$:
\be\label{eq:cut}\int D\dphi\;e^{-\frac{1}{2\hbar}\int d^4x\,\dphi\calO\dphi}\,F[\dphi(0,\cdot)]&=&\int D\chi\;\Psi_-[\chi]\,\Psi_+[\chi]\,F[\chi],\ee
where $\Psi_\mp[\chi]$ are the half-space integrals with Dirichlet data $\chi$. By the computation of Sec.~\ref{ssec:wf},
$\Psi_\mp[\chi]\propto\exp\left(-\frac{1}{2\hbar}\int d^3x\,\chi M_\mp\chi\right),$ 
with $M_-=M$ and $M_+\chi=-\partial_\tau\eta_+(0^+,\bx)$ defined by the solution $\eta_+$ in $\tau>0$. Since $\phi_b$ is even in $\tau$, $\eta_+(\tau,\bx)=\eta(-\tau,\bx)$ and $M_+=M_-=M$. Hence, under the tunneling state
\be\label{eq:glue}\VEV{F}&=&\frac{\int D\chi\;e^{-\frac{1}{\hbar}\int d^3x\,\chi M\chi}\,F[\chi]}{\int D\chi\;e^{-\frac{1}{\hbar}\int d^3x\,\chi M\chi}}.\ee

Let $G_E(x,y)$ be the full-space Euclidean Green's function of $\calO$ at a fixed nucleation point. The four translation zero modes $\psi_\mu=\partial_\mu\phi_b/\sqrt{S_b}$ 
are the collective coordinates of the nucleation point, 
and $G_E$ is the inverse of $\calO$ on their complement,
\be\label{eq:Gdef}\VEV{\dphi(x)\dphi(y)}&=&\hbar\,G_E(x,y),\no\\
\calO_xG_E(x,y)&=&\delta^4(x-y)-\sum_\mu\psi_\mu(x)\psi_\mu(y),\;\;\;\;\;\int d^4x\,\psi_\mu(x)G_E(x,y)=0.\ee
%
Taking $F=\chi(\bx)\chi(\by)$ in \refeq{glue} and comparing with \refeq{cov},
\be\label{eq:GM}\VEV{\chi(\bx)\chi(\by)}_{\Psi_T}&=&\frac{\hbar}{2}M^{-1}(\bx,\by)\;=\;\hbar\,G_E\big((0,\bx),(0,\by)\big).\ee
Thus the equal-time covariance of the field in the tunneling state is the Euclidean propagator in the bounce background evaluated on the nucleation slice. 

Eq.~(\ref{eq:GM}) supports the intuitive probabilistic reading of Callan \& Coleman's fluctuation integral \refeq{gauss}, discussed briefly at the end of Sec.~\ref{ss:bounce}. 
To see this, expand the fluctuation in the orthonormal eigenmodes of $\calO$, $\dphi(x)=\sum_nc_n\varphi_n(x)$, $\calO\varphi_n=\lambda_n\varphi_n$, $\int d^4x\,\varphi_n\varphi_m=\delta_{nm}$. On the complement of the zero modes, the Green's function has the spectral decomposition
\be\label{eq:modesum}G_E\big((0,\bx),(0,\by)\big)&=&\sum_{\lambda_n\neq0}\frac{\varphi_n(0,\bx)\varphi_n(0,\by)}{\lambda_n}.\ee
Now, referring back to the Gaussian integral \refeq{gauss}, up to the ambiguous negative mode it is $\prod_n\int dc_n\,e^{-\lambda_nc_n^2/2\hbar}$, which invites the interpretation that the probability for the bubble to nucleate decorated by mode $n$ with amplitude $c_n$ is $\propto e^{-\lambda_nc_n^2/2\hbar}$, so that 
\be\label{eq:CCcomp}\VEV{c_nc_m}&=&\frac{\prod_k\int dc_k\,e^{-\lambda_kc_k^2/2\hbar}c_nc_m}{\prod_k\int dc_k\,e^{-\lambda_kc_k^2/2\hbar}}\,=\,\hbar\frac{\delta_{nm}}{\lambda_n}.\ee 
This is precisely \refeq{GM}, via the representation \refeq{modesum}. Note that the sum in \refeq{modesum} contains the negative mode with its {\it signed} eigenvalue, and \refeq{CCcomp} reproduces it along the rotated contour of \refeq{gauss}: $\VEV{c_-^2}=\hbar/\lambda_-<0$, resolving the ambiguity in the naive probabilistic interpretation.

A few more comments are in order, starting with the zero modes of $\calO$. The spatial translations $\partial_i\phi_b$ are slice data, and exact zero modes of $M$. They shift the bubble as a whole: since we compute the statistics of fluctuations around a bubble at a given nucleation point, we ignore these zero modes. We will specifically focus our attention on fluctuations and observables that are orthogonal to these zero modes. 

The time translation $\partial_\tau\phi_b$ vanishes on the slice, so it is not slice data. It is also not an exact zero mode of the half-space action. It is a zero mode of the half-space quadratic action, however, and that allows us to exhibit the domain of $\chi$ over which $M$ is defined. Let $\eta$ be the solution of \refeq{eta} for data $\chi$. Since $\calO\eta=0$, $\calO\partial_\tau\phi_b=0$ and $\partial_\tau\phi_b|_{\tau=0}=0$, integration by parts on the half space gives
\be\label{eq:green}0\,=\,\int_{\tau<0}d^4x\left(\partial_\tau\phi_b\,\calO\eta-\eta\,\calO\partial_\tau\phi_b\right)&=&\int d^3x\,\chi(\bx)\,\partial_\tau^2\phi_b\Big|_{\tau=0}\,=\,\int d^3x\,\chi(\bx)\,\frac{\phi_b'(\rho)}{\rho}.\ee
Hence \refeq{eta} has a solution only for data $\chi$ with
\be\label{eq:esc}\int d^3x\,\chi(\bx)\,\frac{\phi_b'(\rho)}{\rho}&=&0.\ee
Thus, $M$ and the Gaussian \refeq{PsiT} are defined only on this subspace of slice data. Indeed, for $\chi$ that does not satisfy \refeq{esc}, the distribution is not Gaussian, because $S_2$ is not a local minimum among half-space functions with slice value $\chi$. To see this, add $c\,\partial_\tau\phi_b$ to any function on the $\tau<0$ half space with slice value $\chi$. This changes $S_2$ by $c\int d^3x\,\chi\,\phi_b'/\rho$, since $S_2[\partial_\tau\phi_b]=0$ by \refeq{S2eta} with zero data. The response of the action is linear in $c$, and there is no quadratic $c^2$ term. This signals the fact that $\phi_b'/\rho$ lies along the escape path, or growth direction, of the bubble: $\phi_b(\sqrt{\rho^2-t^2})=\phi_b(\rho)-\frac{t^2}{2}\phi_b'(\rho)/\rho+O(t^4)$.

Another way to phrase the admissibility condition, \refeq{esc}, for the slice fluctuations $\chi$ over which the Gaussian probability distribution is defined, is as a definition of our choice of bubble nucleation slice: it is the slice of zero potential energy, equivalent to the classical turning point in the quantum mechanical tunneling analogue~\cite{Coleman:1977py}. Indeed, the configuration space potential energy $U$ and the total Euclidean energy $E$ on any time slice are
\be
\label{eq:energy}U&=&\int d^3x\left[\frac{1}{2}\left(\nabla\phi\right)^2+V(\phi)\right],\,\,\,\,\,E\,=\,\int d^3x\,\frac{1}{2}\left(\partial_\tau\phi\right)^2-U.
\ee
Evaluated on the bounce, \refeq{bounce} keeps $E=0$ from the false-vacuum to the nucleation point, while $U$ starts out zero, grows to form the barrier, and vanishes again at the nucleation point. If we venture slightly off the bounce, perturbing at time $\tau$ and using \refeq{bounce} gives 
\be\delta U&=&\int d^3x\,\delta\phi(\tau,{\bf x})\left[-\nabla^2\phi_b+V'(\phi_b)\right]\,=\,\int d^3x\,\delta\phi(\tau,{\bf x})\,\partial_\tau^2\phi_b(\tau,{\bf x}).\ee
On the nucleation slice, demanding $\delta U=0$ amounts to \refeq{esc}.

For our goal of computing the quadrupole moment of the nucleating bubble, the fluctuation subspace that satisfies \refeq{esc} is sufficient: since the function $\phi_b'/\rho$ is spherically symmetric, the quadrupole moment is built solely of fluctuations orthogonal to it.


\section{Root-mean-square quadrupole moment}\label{sec:quad}
On the slice $t=0$ the momentum of the background vanishes, $\pi_b=\partial_t\phi_b|_{t=0}=i\partial_\tau\phi_b|_{\tau=0}=0$, so the full momentum there is the fluctuation, $\pi=\dpi$. The energy density $T_{00}=\frac{1}{2}\pi^2+\frac{1}{2}(\nabla\phi)^2+V(\phi)$ can be expanded in powers of the fluctuation fields $(\dphi,\dpi)$ as
\be\label{eq:T00}T_{00}&=&T^{b}_{00}+T^{(1)}+T^{(2)}+\dots,\vs
\label{eq:T1}T^{(1)}&=&\nabla\phi_b\cdot\nabla\dphi+V'(\phi_b)\,\dphi,\vs
\label{eq:T2}T^{(2)}&=&\frac{1}{2}\dpi^2+\frac{1}{2}\left(\nabla\dphi\right)^2+\frac{1}{2}V''(\phi_b)\,\dphi^2.\ee
The quadrupole moment of the energy density with respect to the nucleation point is
\be\label{eq:Qdefop}Q_{ij}&=&\int d^3x\left(x_ix_j\right)^\TL\,T_{00}(0,\bx),\;\;\;\;\;\left(x_ix_j\right)^\TL\,\equiv\,x_ix_j-\frac{1}{3}\delta_{ij}\rho^2.\ee
The zeroth-order (bounce) contribution vanishes, $\int(x_ix_j)^\TL T^b_{00}=0$, because the bounce is spherical, and the expectation value $\VEV{Q_{ij}}$ also vanishes because the state \refeq{PsiT} has no preferred direction. The variance starts at $O(\hbar)$  from $\VEV{T^{(1)}T^{(1)}}$. 
Since $T^{(1)}$ does not contain the momentum $\delta\pi$, only the field covariance \refeq{GM} is needed. 

To bring $T^{(1)}$ to a more convenient form, integrate the gradient term by parts. With $\nabla\phi_b=\phi_b'(\rho)\hat n$, $\nabla\cdot(\phi_b'\hat n)=\phi_b''+2\phi_b'/\rho$, and $\hat n\cdot\nabla(x_ix_j)^\TL=(2/\rho)(x_ix_j)^\TL$, we have 
\be Q^{(1)}_{ij}&=&\int d^3x\left(x_ix_j\right)^\TL\left[V'(\phi_b)-\phi_b''-\frac{4}{\rho}\phi_b'\right]\dphi\,=\,-\int d^3x\left(x_ix_j\right)^\TL\frac{\phi_b'(\rho)}{\rho}\,\dphi(\bx),\ee
where the second step uses the bounce equation \refeq{bounce} at $r=\rho$. 
Thus, to leading order,
\be\label{eq:Qlin}Q_{ij}&=&-\int_0^\infty d\rho\,\rho^3\phi_b'(\rho)\int d\Omega\left(\hat n_i\hat n_j\right)^\TL\dphi(\rho\hat n).\ee

\subsection{Angular decomposition}\label{ssec:ang}
Because $\calO$ is $O(4)$-symmetric, its eigenmodes separate in hyperspherical harmonics $\mathcal{Y}_{JLM}(\hat x)$ on $S^3$, with $J=0,1,2,\dots$, $L=0,\dots,J$, $M=-L,\dots,L$, degeneracy $(J+1)^2$ and $-\nabla^2_{S^3}\mathcal{Y}_{JLM}=J(J+2)\mathcal{Y}_{JLM}$; see~\refapp{hsh}. $L$ and $M$ are the quantum numbers of the rotations about the nucleation point. In the sector $J$ the fluctuation operator is
\be\label{eq:OJ}\calO_J&=&-\frac{d^2}{dr^2}-\frac{3}{r}\frac{d}{dr}+\frac{J(J+2)}{r^2}+V''(\phi_b(r)).\ee
Parametrizing $\hat x=(\cos\beta,\sin\beta\,\hat n)$ with $\beta$ the polar angle from the $\tau$ axis, the harmonics are
\be\label{eq:YJLM}\mathcal{Y}_{JLM}(\hat x)&=&N_{JL}(\beta)\,Y_{LM}(\hat n),\;\;\;\;\;\;N_{JL}(\beta)\,=\,2^LL!\sqrt{\frac{2(J+1)(J-L)!}{\pi(J+L+1)!}}\;C^{L+1}_{J-L}(\cos\beta)\,\sin^L\!\beta,\ee
with $C^\lambda_n$ the Gegenbauer polynomials. 
On the slice, $\beta=\pi/2$, and $N_{JL}(\pi/2)$ vanishes for $J-L$ odd because $C^{L+1}_{J-L}(0)$ does. 
Under the reflection $\tau\to-\tau$, i.e., $\beta\to\pi-\beta$, $\mathcal{Y}_{JLM}\to(-1)^{J-L}\mathcal{Y}_{JLM}$. Hence the fluctuations that survive on the nucleation slice belong to the $\tau$-even angular sectors with
\be\label{eq:tower}J&=&L,\;L+2,\;L+4,\;\dots\,.\ee
Fluctuations with $J-L$ odd are $\tau$-odd, and vanish on the slice. These fluctuations carry the normal derivative, that is, the slice momentum. 

The negative fluctuation eigenmode belongs to the sector 
$(J,L)=(0,0)$. The time-translation zero mode $\partial_\tau\phi_b=\cos\beta\,\phi_b'$ belongs to the $\tau$-odd harmonic $(J,L)=(1,0)$, vanishes on the slice, and is a zero mode of the Dirichlet problem \refeq{eta} rather than of $M$. The spatial translations $\partial_i\phi_b=\sin\beta\,\hat n_i\phi_b'$ are $\tau$-even harmonics in $(J,L)=(1,1)$ and are slice data. For every $J\geq2$, $\calO_J$ is a positive operator: the centrifugal term gives $\calO_J\geq\calO_1$ for $J\geq1$, and $\calO_1\geq0$ because its ground state is the nodeless $\phi_b'$, at eigenvalue zero (Sec.~\ref{ssec:conf}). Importantly for us in this paper, the quadrupole sector $L=2$ modes that survive on the slice contain only $J=2,4,6,\dots$. 

The harmonic decomposition of $G_E$ is
\be\label{eq:GEdec}G_E(x,y)&=&\sum_{J=0}^\infty G_J(r,r')\,\frac{J+1}{2\pi^2}\,U_J(\cos\gamma),\;\;\;\;\cos\gamma=\hat x\cdot\hat y,\ee
where $U_J$ is the Chebyshev polynomial of the second kind, $U_J(\cos\gamma)=\sin((J+1)\gamma)/\sin\gamma$, and $G_J$ is the radial Green's function of \refeq{OJ},
\be\label{eq:GJdef}\calO_J\,G_J(r,r')&=&\frac{\delta(r-r')}{r^3}.\ee
For two points on the slice, $\cos\gamma=\hat n\cdot\hat n'$ is the cosine of the ordinary three-dimensional angle between them, and the angular dependence can be expanded in Legendre polynomials,
\be\label{eq:UP}U_J(c)&=&\sum_{L}u_{JL}\,P_L(c),\;\;\;\;\;u_{JL}\,=\,\frac{2L+1}{2}\int_{-1}^1dc\,U_J(c)P_L(c).\ee
The coefficients $u_{JL}$ are nonzero only for $L\leq J$ and $J-L$ even, consistent with \refeq{tower}. Defining the moments of the slice field $\chi=\dphi(0,\cdot)$ in the usual way,
\be\label{eq:dphiLM}\dphi_{LM}(\rho)&=&\int d\Omega\,Y^*_{LM}(\hat n)\dphi(0,\rho\hat n),
\ee
we can write the sector $L$ contribution to \refeq{GM} as
%
\be\label{eq:gL}\VEV{\dphi_{LM}(\rho)\dphi^*_{L'M'}(\rho')}&=&\hbar\,\delta_{LL'}\delta_{MM'}\sum_{J\geq L,J-L~{\rm even}}c^{(L)}_J\,G_J(\rho,\rho'),\ee
%
where 
\be\label{eq:cLJ} c^{(L)}_J\,=\,\frac{2(J+1)\,u_{JL}}{\pi(2L+1)}.\ee

\subsection{The $L=2$ covariance}
To compute the RMS of the quadrupole \refeq{Qlin}, we need \refeq{gL} only at $L=2$,  
where the coefficients of Eq.~(\ref{eq:cLJ}) have the closed form (\refapp{hsh})
\be\label{eq:cJ}c_J\equiv c^{(2)}_J&=&\frac{2}{\pi}\,\frac{J(J+2)}{(J-1)(J+3)},\;\;\;\;\;J=2,4,6,\dots,\ee
with $c_2=16/5\pi$ and $c_J\to2/\pi$ at large $J$. 
Note that the kernel $(\hat n_i\hat n_j)^\TL$ is a combination of the $Y_{2M}$s, 
\be&&\sum_M\int d\Omega\,(\hat n_i\hat n_j)^\TL Y_{2M}(\hat n)\int d\Omega'\,(\hat n'_k\hat n'_l)^\TL Y^*_{2M}(\hat n')\no\\
&&\qquad=\,\int d\Omega\,(\hat n_i\hat n_j)^\TL(\hat n_k\hat n_l)^\TL\,=\,\frac{4\pi}{15}\left(\delta_{ik}\delta_{jl}+\delta_{il}\delta_{jk}-\frac{2}{3}\delta_{ij}\delta_{kl}\right).\ee
With this, we obtain the quadrupole variance:
\be\label{eq:QQ}\VEV{Q_{ij}Q_{kl}}&=&\hbar\frac{4\pi}{15}\left(\delta_{ik}\delta_{jl}+\delta_{il}\delta_{jk}-\frac{2}{3}\delta_{ij}\delta_{kl}\right)\calQ^2,\\
\label{eq:Qdef0}\calQ^2&=&\sum_{J=2,4,\dots}\calQ_J^2,\\
\label{eq:Qdef}\calQ^2_J&=&c_J\int_0^\infty d\rho\int_0^\infty d\tilde\rho\,\rho^3\phi_b'(\rho)\,\tilde\rho^3\phi_b'(\tilde\rho)\,G_J(\rho,\tilde\rho).\ee
This gives, for example,
\be\label{eq:dR2}\VEV{Q_{12}^2}\,=\,\hbar\frac{4\pi}{15}\calQ^2,\;\;\;\;\;\sum_{ij}\VEV{Q_{ij}Q_{ij}}\,=\,\hbar\frac{8\pi}{3}\calQ^2.\ee
%

\subsection{Short-distance structure}\label{ssec:UV}
At short distance $G_E(x,y)\to1/(4\pi^2|x-y|^2)$, so the equal-time kernel in \refeq{Qdef} behaves as $1/|\bx-\by|^2$, which is integrable in $d^3y$. For a bounce whose wall, at radius $R$, has a finite width $\ell_w$, the two radial integrals in \refeq{Qdef} therefore converge at short distance (at large distance they converge when the false vacuum is massive; for a massless false vacuum they diverge logarithmically, Sec.~\ref{ssec:lightfv}), and so does the sum over $J$: for $J\gg R/\ell_w$ the centrifugal term dominates $\calO_J$, $G_J(\rho,\rho')\simeq(\rho_</\rho_>)^J/[2(J+1)\rho_>^2]$ is confined to $|\rho-\rho'|\lesssim\rho/J$, where $\int d\rho'\,G_J(\rho,\rho')f(\rho')\to f(\rho)/[(J+1)^2\rho]$ for smooth $f$, and
\be\label{eq:tail}\calQ^2_J\;\to\;c_J\,\frac{1}{(J+1)^2}\int_0^\infty d\rho\,\rho^5\phi_b'(\rho)^2\;\;\;\;\;\;(J\gg R/\ell_w).\ee
The limit of a wall of zero thickness, $\phi_b'\to-\Delta\phi\,\delta(\rho-R)$ with $\Delta\phi$ the field excursion across the wall, cannot be taken inside the sum over $J$: it would replace the double integral by $\Delta\phi^2R^6G_J(R,R)$, whose sum over $J$ diverges logarithmically because of the coincident-point singularity of $G_E$. The thin-wall limit of the result is nevertheless regular (Sec.~\ref{ssec:thin}).

\subsection{A numerics-friendly version of the radial Green's function}\label{ssec:GJ}
%
We can build $G_J$ from two homogeneous solutions of $\calO_Ju=0$: $u^<_J$, satisfying $u^<_J\propto r^J$ and regular at $r\to 0$, and $u^>_J$, satisfying $u^>_J\propto K_{J+1}(mr)/r$ and decaying at $r\to\infty$. For $J\geq2$ neither solution has a zero, since $\calO_J$ is positive, and we have 
\be\label{eq:GJ}G_J(r,r')&=&\frac{u^<_J(r_<)\,u^>_J(r_>)}{\mathcal{W}_J},\;\;\;\;\;\mathcal{W}_J\,=\,r^3\left[\frac{du^<_J}{dr}\,u^>_J(r)-u^<_J(r)\,\frac{du^>_J}{dr}\right]\,>\,0,\ee
where $r_<$ ($r_>$) is the smaller (larger) of $r,r'$ and the Wronskian $\mathcal{W}_J$ is independent of $r$. 

The logarithmic derivatives $w^{<,>}_J=d\ln u^{<,>}_J/dr$ obey the Riccati equation
\be\label{eq:riccati}\frac{dw}{dr}&=&-w^2-\frac{3}{r}w+\frac{J(J+2)}{r^2}+V''(\phi_b(r)),\ee
with $w^<_J=J/r+O(r)$ at $r\to0$ and $w^>_J=-mK_J(mr)/K_{J+1}(mr)-(J+2)/r$ at $r\to\infty$. This allows us to cast $G_J$ in a form that is convenient for numerical integration,
\be\label{eq:GJw}G_J(r_<,r_>)&=&\frac{\exp\left[-\int_{r_<}^{r_>}dr\,w^<_J(r)\right]}{r_>^3\left[w^<_J(r_>)-w^>_J(r_>)\right]}.\ee
We use this presentation of $G_J$ for numerical calculations.

\subsection{The wall-displacement mode}\label{ssec:conf}
%
Defining $\eta=\ln r$, we have, for any function $u(r)$,
\be\label{eq:conf}\calO_J\,u&=&r^{-3}\left[\mathcal{K}+(J+1)^2\right](r\,u),\;\;\;\;\;\;\mathcal{K}\,\equiv\,-\frac{d^2}{d\eta^2}+r^2V''(\phi_b(r)).\ee
Note that the operator $\mathcal{K}$ makes no reference to $J$. Let $\{F_n\}$ be the complete set of eigenfunctions of $\mathcal{K}$, bound states and continuum,
\be\label{eq:Kspec}\mathcal{K}F_n&=&\kappa_n\,F_n,\;\;\;\;\;\;\int d\eta\,F_nF_m\,=\,\delta_{nm},\ee
with Dirac-delta normalization understood for the continuum. By \refeq{conf}, the radial Green's function of \refeq{GJdef} is
\be\label{eq:GJspec}G_J(r,r')&=&\sum_n\frac{F_n(\eta)F_n(\eta')}{r\,r'\left[\kappa_n+(J+1)^2\right]},\ee
as one can check by applying $\calO_J$ and using $\sum_nF_n(\eta)F_n(\eta')=\delta(\eta-\eta')=r\,\delta(r-r')$.

One eigenfunction can be found immediately in closed form: the translations $\partial_\mu\phi_b=(x_\mu/r)\phi_b'$ are zero modes of $\calO$ in the sector $J=1$, so $\calO_1\phi_b'=0$, and \refeq{conf} at $J=1$ gives
\be\label{eq:F0}F_0(\eta)&=&r\,\phi_b'(r),\;\;\;\;\;\;\kappa_0\,=\,-4.\ee
$F_0$ has no node, because the bounce profile is monotonic\footnote{We assume the minimum-action $O(4)$-symmetric bounce and do not consider subleading saddles~\cite{Avraham:2026mgk}.}, $\phi_b'<0$ for $r>0$,
so it is the ground state of $\mathcal{K}$: $\kappa_n>-4$ for $n\geq1$. We call $F_0$ the {\it wall-displacement mode}, because a slice field $\dphi\propto\phi_b'(\rho)$ corresponds to a rigid radial displacement of the profile.

Notice that the integration weight in \refeq{Qdef} is $\rho^3\phi_b'=\rho^2F_0(\eta)$. With $d\rho=\rho\,d\eta$, and using \refeq{GJspec}, we have
%
\be\label{eq:Qspec}\calQ^2_J&=&c_J\,N_0\sum_n\frac{\left[\int d\eta\,\rho^2\,\hat F_0F_n\right]^2}{\kappa_n+(J+1)^2},\;\;\;\;\hat F_0\equiv\frac{F_0}{\sqrt{N_0}},\;\;\;\;N_0=\int d\eta\,F_0^2=\int_0^\infty d\rho\,\rho\,\phi_b'(\rho)^2.\ee
%
For $J\geq2$, every term in the sum is positive, with $\kappa_n+(J+1)^2\geq\kappa_0+9=5$. The $n=0$ term involves
\be\label{eq:virial}\Sigma_3\,\equiv\,\int d\eta\,\rho^2F_0^2\,=\,\int_0^\infty d\rho\,\rho^3\phi_b'(\rho)^2\,=\,\frac{2}{\pi^2}\,S_b,\ee
where the last equality comes from the bounce equation\footnote{This result reflects the stationarity of the action under dilatations $\phi_b(r)\to\phi_b(r/a)$: $\int d^4x\,V(\phi_b)=-\frac{1}{4}\int d^4x\,(\partial\phi_b)^2$, so that $S_b=\frac{1}{4}\int d^4x\,(\partial\phi_b)^2=\frac{\pi^2}{2}\int dr\,r^3\phi_b'^2$.}. 
%

If we keep only the $n=0$ term of \refeq{Qspec}, we can perform the sum over $J$ in \refeq{Qdef0}. Let us denote the result -- the value of $\calQ^2$ obtained restricting to the $n=0$ mode, but summing over $J$ -- by $\calQ^2_{\rm wall}$; we will see that for some models, $\calQ^2_{\rm wall}$ is a reasonable approximation of the full $\calQ^2$. We find 
\be\label{eq:Qw}\calQ^2_{\rm wall}&=&\frac{\Sigma_3^2}{N_0}\sum_{J=2,4,\dots}\frac{c_J}{J(J+2)-3}\;=\;\frac{3\pi}{32}\,\frac{\Sigma_3^2}{N_0},\ee
where we used 
%
$\sum_{J=2,4,\dots}\frac{c_J}{(J-1)(J+3)}=\frac{2}{\pi}\sum_{k=1}^\infty\frac{4k(k+1)}{(2k-1)^2(2k+3)^2}
=\frac{3\pi}{32}.$ 

Using the orthogonality of the $F_n$s, the terms $n\geq1$ can be written as $\int d\eta\,\rho^2\hat F_0F_n=\int d\eta\,(\rho^2-\Sigma_3/N_0)\hat F_0F_n$. Thus the modes beyond the wall displacement couple to the quadrupole only through the variation of $\rho^2$ across the weight $\hat F_0^2$, whose mean is $\Sigma_3/N_0$. We call their contribution the remainder, $\calQ^2_{\rm rem}\equiv\calQ^2-\calQ^2_{\rm wall}$. The remainder tends to be small in the thin-wall limit. 

\section{Examples}\label{sec:model}
\subsection{Thin wall}\label{ssec:thin}
When the difference $\epsilon$ between the energy densities of the false and true vacua is small compared with the barrier, the bounce is a wall of tension $\sigma=\int d\rho\,\phi_b'^2$ and width $\ell_w$ at radius $R=3\sigma/\epsilon\gg\ell_w$~\cite{Coleman:1977py}. The moments of the profile reduce to
\be\label{eq:thinmom}\Sigma_3\to\sigma R^3,\;\;\;\;\;\;N_0\to\sigma R,\;\;\;\;\;\;S_b\to\frac{\pi^2}{2}\,\sigma R^3,\ee
and the quadrupole variance is, choosing the $1,2$ entry for example,
\be\VEV{Q_{12}^2}&\approx&\hbar\frac{4\pi}{15}\calQ^2_{\rm wall}\,=\,\hbar\frac{\pi^2}{40}\sigma R^5\;\;\;\;\;\;(\text{thin wall}).\ee

\subsubsection{Comparison with the analysis of~\cite{Blum:2024hcs}}\label{ssec:bnb}
We can compare the thin-wall result with the analysis of Ref.~\cite{Blum:2024hcs}. For a wall at $R(\hat n)=R+\delta R(\hat n)$ the energy density is $\sigma\delta(\rho-R(\hat n))-\epsilon\,\Theta(R(\hat n)-\rho)$, and \refeq{Qdefop} gives, to first order in $\delta R$, $Q_{ij}=(4\sigma R^3-\epsilon R^4)\int d\Omega\,(\hat n_i\hat n_j)^\TL\delta R=\sigma R^3\int d\Omega\,(\hat n_i\hat n_j)^\TL\delta R(\hat n)$. \refeq{Qlin} gives the same with $\dphi=-\phi_b'\,\delta R(\hat n)$, the field of the displaced wall, recalling $\Sigma_3\to\sigma R^3$. Expanding $\delta R(\hat n)=\sum_{LM}\delta R_{LM}Y_{LM}(\hat n)$, only $L=2$ enters, and $\sum_{ij}Q_{ij}Q_{ij}=\sigma^2R^6\,\frac{8\pi}{15}\sum_M|\delta R_{2M}|^2$, using $\int d\Omega\,d\Omega'\,[(\hat n\cdot\hat n')^2-\frac{1}{3}]Y_{2M}(\hat n)Y^*_{2M'}(\hat n')=\frac{8\pi}{15}\delta_{MM'}$. Comparing with \refeq{dR2}, and noting that the sum on $M$ runs from $-2$ to $2$, we have (for any of the $M$s)
\be\label{eq:dR2tw}\hbar\calQ^2&=&\sigma^2R^6\VEV{|\delta R_{2M}|^2}\;\;\;\;\;\;(\text{thin wall}).\ee
Approximating $\calQ^2$ by \refeq{Qw}, and using \refeq{thinmom}, we find
\be\label{eq:dRw}\VEV{|\delta R_{2M}|^2}&=&\hbar\frac{\calQ^2_{\rm wall}}{\sigma^2R^6}\,=\,\frac{3\pi}{32}\,\frac{\hbar}{\sigma R}.\ee

\refeq{dRw} reproduces the variance at $t=0$ of the wall-displacement mode in the state used in~\cite{Blum:2024hcs}. There, the wall is described by the Nambu--Goto action on its world volume, the displacement $\zeta=\delta R/R$ obeys $[\Box_{{\rm dS}_3}+3]\zeta=0$ on the unit ${\rm dS}_3$ into which the Euclidean $S^3$ continues, and the state is fixed by requiring the mode functions to be regular on the $S^3$ (the Hartle--Hawking prescription~\cite{Hartle:1983ai}, equivalent to the Bunch--Davies state on the ${\rm dS}_3$ leaves of Refs.~\cite{Garriga:1991tb,Garriga:1996pg}). The resulting mode functions are $f_L(\tau)=\frac{N_L}{4}[2Q_{L,2}(\tanh\tau)+i\pi P_{L,2}(\tanh\tau)]/\cosh\tau$ with $N_L^2=4\Gamma(L-1)/[\pi\sigma R^3\Gamma(L+3)]$ (Eqs.~(10)--(11) of Ref.~\cite{Blum:2024hcs}, with $l$ there our $L$). At $\tau=0$ and $L=2$, $Q_{2,2}(0)=0$ and $P_{2,2}(0)=3$, so that $\VEV{|\zeta_{2M}|^2}=|f_2(0)|^2=3\pi/(32\sigma R^3)$, which is \refeq{dRw}. The regularity of the mode functions on the Euclidean $S^3$ is the world-volume version of the half-bounce boundary condition $\eta\to0$ at $\tau\to-\infty$ in \refeq{eta}, so the Hartle--Hawking state of the wall mode is the tunneling state \refeq{PsiT} restricted to the ground state $F_0$ of $\mathcal{K}$. Our full-field calculation is then consistent with the $n=0$ term comprising the previous analysis~\cite{Blum:2024hcs}, but allows us to include the rest of the spectrum of $\mathcal{K}$.

\subsubsection{Internal excitations of a thin wall}\label{ssec:rest}
The structure of the remainder $\calQ^2_{\rm rem}$ in the thin-wall limit follows from the orthogonality argument of Sec.~\ref{ssec:conf}. Near the wall, $\eta=\ln R+y/R$ with $y=\rho-R$, the operator $\mathcal{K}$ becomes $R^2[-d^2/dy^2+V''(\phi_b)]$. Its spectrum above $F_0$ consists of the transverse excitations of the planar wall, $\kappa_n=\omega_n^2R^2$ with $\omega_n$ the frequencies of the one-dimensional problem $[-d^2/dy^2+V''(\phi_b(y))]f_n=\omega_n^2f_n$, and $F_n=\sqrt R\,f_n(y)$ with $\int dy\,f_n^2=1$.\footnote{The spectrum of $\mathcal{K}$ also contains modes supported in the interior of the bubble, with $\kappa_n\geq0$; their amplitude at the wall is exponentially small and we neglect them.} Expanding $\rho^2-R^2=2Ry+y^2$, the overlaps in \refeq{Qspec} are $2R\VEV{y}_{0n}+\VEV{y^2}_{0n}$ with $\VEV{y^k}_{0n}=\int dy\,y^kf_0f_n$. 
For $J\ll\omega_nR$ the denominators are $\omega_n^2R^2$, and the sum over even $J$ of $c_J/[\omega_n^2R^2+(J+1)^2]$ is $1/(2\omega_nR)$ to leading order in $1/\omega_nR$. Hence
\be\label{eq:Qrest}\calQ^2_{\rm rem}&\to&2\,\sigma R^2K,\;\;\;\;\;\;K\,\equiv\,\sum_{n\geq1}\frac{\VEV{y}_{0n}^2}{\omega_n}\;\;\;\;\;\;(\text{thin wall}),\ee
using $N_0\to\sigma R$. $K$ (of dimension length$^3$) is a property of the planar wall alone. 
The ratio of the two parts is
\be\label{eq:ratio}\frac{\calQ^2_{\rm rem}}{\calQ^2_{\rm wall}}&\to&\frac{64K}{3\pi}\,\frac{1}{R^3}\;\;\;\;\;\;(\text{thin wall}).\ee
For a wall with a single internal scale, $K$ is set by the internal spectrum of the wall and should scale as $K\sim\ell_w^3$. Thus, $\calQ^2_{\rm rem}$ is suppressed relative to $\calQ^2_{\rm wall}$ by the third power of the ratio of wall width to bubble radius. 

For thick walls the wall-displacement mode contribution $\calQ^2_{\rm wall}$ can become subdominant to the internal modes $\calQ^2_{\rm rem}$. When the false vacuum is light, $mR\lesssim1$, the bounce has a power-law tail $\phi_b-\phi_{\rm FV}\propto\rho^{-2}$ out to $\rho\sim1/m$, over which the weight $\rho^3\phi_b'$ is constant rather than localized, and the modes supported on the tail couple to the quadrupole through $\rho^2-R^2\simeq\rho^2$, not through the curvature of the wall. We analyze such examples in Secs.~\ref{ssec:lightfv} and~\ref{ssec:lfvnum}.

\subsection{One-parameter quartic potentials}\label{ssec:family}
Consider the one-parameter family of potentials
\be\label{eq:U}V(\phi)&=&\frac{m^4}{g}\,U(x),\;\;\;\;\;x\,=\,\frac{\sqrt g\,\phi}{m},\;\;\;\;\;U(x)\,=\,\frac{1}{2}x^2-\frac{1}{2}x^3+\frac{\alpha}{8}x^4,\;\;\;\;\;0\leq\alpha<1.\ee
The false vacuum is at $x=0$ with mass $m$. The barrier top and the true vacuum are at $x=(3/2\mp\sqrt{9/4-2\alpha})/\alpha$. The vacua become degenerate at $\alpha=1$, where $x_{\rm TV}=2$ and $U=\frac{1}{2}x^2(1-x/2)^2$, and the true vacuum runs to infinity at $\alpha\to0$. The action is proportional to $1/g$, $S_b=\hat S(\alpha)/g$, so the dimensionless coupling $g$ is the semiclassical expansion parameter. $\alpha$ controls the shape of the bounce from a thick wall at small $\alpha$ to a thin wall at $\alpha\to1$. In units of $m$ and with $s=mr$, the bounce is $\phi_b=(m/\sqrt g)\,x_b(s)$ with $x_b''+(3/s)x_b'=U'(x_b)$, and the relevant quantities entering the quadrupole analysis scale as
\be\label{eq:scalings}S_b=\frac{\hat S(\alpha)}{g},\;\;\;\;\sigma=\frac{m^3\hat\sigma}{g},\;\;\;\;\Sigma_3=\frac{\hat\Sigma_3}{g},\;\;\;\;N_0=\frac{m^2\hat N_0}{g},\;\;\;\;\calQ^2=\frac{q^2(\alpha)}{g\,m^2},\ee
with $\hat S(\alpha)=2\pi^2\int ds\,s^3[\frac{1}{2}x_b'^2+U(x_b)]$, $\hat\sigma=\int ds\,x_b'^2$, $\hat\Sigma_3=\int ds\,s^3x_b'^2$, $\hat N_0=\int ds\,s\,x_b'^2$, and
\be\label{eq:q}q^2(\alpha)&=&\sum_{J=2,4,\dots}c_J\int_0^\infty ds\int_0^\infty ds'\,s^3x_b'(s)\,s'^3x_b'(s')\,\hat G_J(s,s'),\ee
$\hat G_J$ being the Green's function of the dimensionless problem. We define the wall position $R$ as the location of the maximum of $|x_b'|$, and the wall width $\ell_w$ as the full width at half maximum of $|x_b'|$. In the thin-wall limit $x_b'\to-\frac{1}{2}{\rm sech}^2[(s-mR)/2]$, so that $m\ell_w\to4\,{\rm arccosh}\sqrt2=3.53$, $\hat\sigma\to2/3$, $mR\to1/(1-\alpha)$ and $\hat S\to\pi^2/[3(1-\alpha)^3]$.

The planar wall of the family is the kink $x=2/(1+e^{my})$ of $U=\frac{1}{2}x^2(1-x/2)^2$, with $V''=m^2[1-\frac{3}{2}{\rm sech}^2(my/2)]$. This is a P\"oschl--Teller potential: the transverse spectrum has the zero mode $f_0\propto{\rm sech}^2(my/2)$, one odd bound state, the shape mode $f_1\propto{\rm sech}(my/2)\tanh(my/2)$ at $\omega_1^2=\frac{3}{4}m^2$, and a continuum above $m^2$. The shape mode gives $\VEV{y}_{01}=\pi/(2\sqrt2\,m)$ and $\VEV{y}_{01}^2/\omega_1=\pi^2/(4\sqrt3\,m^3)\approx1.42/m^3$, and the odd continuum adds $\approx0.05/m^3$ (numerically), so that \refeq{ratio} becomes
\be\label{eq:K}K&=&\frac{1.47}{m^3},\;\;\;\;\;\;\frac{\calQ^2_{\rm rem}}{\calQ^2_{\rm wall}}\to\frac{10.0}{(mR)^3}.\ee
In this family the wall width, the shape-mode frequency and the false-vacuum mass are all set by $m$ ($m\ell_w=3.53$, $\omega_1=0.87m$), so the thin-wall limit is $mR\gg1$. In terms of the wall width, the suppression \refeq{K} is $0.23\,(\ell_w/R)^3$. 

\reftab{bounce} lists some numerical results for different values of $\alpha$. The thin-wall relations $R=3\sigma/\epsilon$ and $S_b=\pi^2\sigma R^3/2$ hold to $1\%$ and $3\%$ at $\alpha=0.9$ and fail by order one at $\alpha\lesssim0.5$, where $\ell_w\gtrsim R$. The relation $S_b=\pi^2\Sigma_3/2$, \refeq{virial}, holds precisely at every $\alpha$.
\begin{table}[h!]
\centering
\sisetup{table-alignment-mode=none,group-digits=none}
\begin{tabular}{lcSSSSSSS}
$\alpha$ & $x_{\rm TV}$ & {$x_b(0)$} & {$mR$} & {$m\ell_w$} & {$\hat\sigma$} & {$\hat S$} & {$\hat\Sigma_3$} & {$\hat N_0$}\\
\hline
0.00 & $\infty$ & 5.7813 & 0.660 & 1.21 & 16.6676 & 90.86 & 18.41 & 13.99\\
0.30 & 9.282 & 4.8590 & 1.160 & 1.78 & 8.5399 & 155.2 & 31.44 & 11.25\\
0.50 & 5.236 & 4.1387 & 1.930 & 2.29 & 4.8304 & 268.1 & 54.33 & 9.689\\
0.90 & 2.412 & 2.4119 & 10.930 & 3.13 & 1.0815 & 7140 & 1447 & 11.82
\end{tabular}
\caption{Numerical analysis for the family \refeq{U}, that allows to explore thick wall at small $\alpha$ and thin wall at $\alpha\to1$. True-vacuum position, field at the center, wall position and width, the moment $\hat\sigma$ (the wall tension in the thin-wall limit), bounce action and the two moments of the wall profile that enter \refeq{Qspec} and \refeq{virial}. All quantities are in units of $m$ with the coupling $g$ scaled out as in \refeq{scalings}. In the thin-wall limit $mR\to1/(1-\alpha)$, $m\ell_w\to3.53$, $\hat\sigma\to2/3$, $\hat\Sigma_3\to\hat\sigma(mR)^3$ and $\hat N_0\to\hat\sigma\,mR$.}
\label{tab:bounce}
\end{table}

\reftab{results} shows the relative deformation $\sqrt\hbar\,\calQ/(\Sigma_3R)$, which in the thin-wall limit is the RMS displacement of the wall per $L=2$ mode in units of its radius, \refeq{dR2tw}, measured in units of $(\hbar/S_b)^{1/2}$. By \refeq{scalings},
\be\label{eq:eps}e(\alpha)\,\equiv\,\frac{\calQ\sqrt{S_b}}{\Sigma_3R}\,=\,\frac{\sqrt{\hat S(\alpha)\,q^2(\alpha)}}{\hat\Sigma_3\,mR}.\ee
A second dimensionless combination is the quadrupole itself in units of the bubble radius and the bounce action,
\be\label{eq:kappa}\calQ&=&\kappa(\alpha)\,R\sqrt{S_b},\;\;\;\;\;\;\kappa(\alpha)\,=\,\frac{\sqrt{q^2(\alpha)/\hat S(\alpha)}}{mR}.\ee
In the thin-wall limit, \refeq{dRw} with \refeq{thinmom} gives
\be\label{eq:thin}e\to\sqrt{\frac{3\pi^3}{64}}=1.206,\;\;\;\;\;\;\kappa\to\sqrt{\frac{3}{16\pi}}=0.244.\ee

\begin{figure}[t]
\centering
\includegraphics[width=0.55\textwidth]{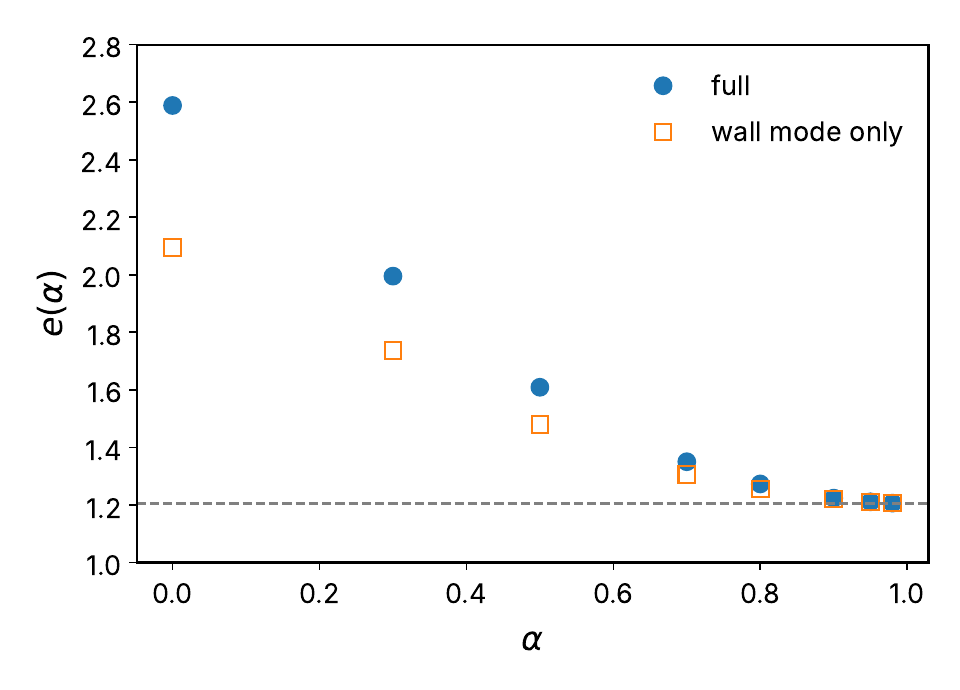}
\caption{The relative $L=2$ deformation per mode in units of $(\hbar/S_b)^{1/2}$, $e(\alpha)$ of \refeq{eps}, from the full result (circles) and from the wall-displacement part alone (squares). The dashed line is the thin-wall limit $1.206$.}
\label{fig:result}
\end{figure}

\reffig{result} shows $e(\alpha)$ against its thin-wall limit \refeq{thin}.

\begin{table}[t]
\centering
\sisetup{table-alignment-mode=none,group-digits=none}\setlength{\tabcolsep}{4pt}
\begin{tabular}{lSSSS[exponent-mode=fixed,fixed-exponent=0]SSSSS}
$\alpha$ & {$q^2_{\rm wall}$} & {$q^2_{\rm rem}$} & {$q^2$} & {$\dfrac{q^2_{\rm rem}}{q^2_{\rm wall}}$} & {$\dfrac{q^2_{\rm rem}}{(mR)^2}$} & {$\dfrac{(mR)^3q^2_{\rm rem}}{q^2_{\rm wall}}$} & {$e(\alpha)$} & {$\kappa(\alpha)$} & {$e_{\rm wall}(\alpha)$}\\[4pt]
\hline
0.00 & 7.137 & 3.752 & 10.89 & 5.26e-01 & 8.614 & 0.15 & 2.588 & 0.5245 & 2.096\\
0.30 & 25.87 & 8.258 & 34.13 & 3.19e-01 & 6.137 & 0.50 & 1.995 & 0.4043 & 1.737\\
0.50 & 89.73 & 16.45 & 106.2 & 1.83e-01 & 4.415 & 1.32 & 1.609 & 0.3261 & 1.479\\
0.90 & 5.218e+04 & 270.1 & 5.245e+04 & 5.18e-03 & 2.261 & 6.76 & 1.224 & 0.2480 & 1.221

\end{tabular}
\caption{The quadrupole variance for the family \refeq{U}. Wall-displacement part $q^2_{\rm wall}$ of \refeq{Qw}, remainder $q^2_{\rm rem}$, total $q^2$ of \refeq{q}, all in the units of \refeq{scalings}; the ratio of the two parts; the remainder and the ratio in the thin-wall normalizations of \refeqs{Qrest}{K}, whose limits are $2\hat\sigma K m^3=1.966$ and $10.0$; and the two dimensionless results $e(\alpha)$ and $\kappa(\alpha)$ of \refeqs{eps}{kappa}. The last column, $e_{\rm wall}$, is $e(\alpha)$ computed from the wall-displacement part alone. Thin-wall limits: $e\to1.206$, $\kappa\to0.244$.}
\label{tab:results}
\end{table}

\subsection{Fubini bounce}\label{ssec:lightfv}
Next we consider a false vacuum whose mass is small compared with the inverse bubble radius, $mR\ll1$. 
To this end consider the scale-invariant negative-$\phi^4$ potential, whose bounce is the Fubini solution~\cite{Fubini:1976jm,Lee:1985uv} with radius parameter $R$,
\be\label{eq:Fub}V=-\frac{\lambda}{4}\phi^4,\;\;\;\;\;\;\phi_b=\sqrt{\frac{8}{\lambda}}\,\frac{R}{\rho^2+R^2},\;\;\;\;\;\;S_b=\frac{8\pi^2}{3\lambda}.\ee
In pure $\phi^4$ theory $R$ is a free parameter. A stabilizing term at large $\phi$, or the running of $\lambda$, can select $R$ without changing the profile much. The Standard Model Higgs potential at large field values is the canonical example~\cite{Isidori:2001bm,Andreassen:2016cvx}, with $R$ set by the scale at which $\lambda$ is most negative and $m_hR\sim10^{-15}$. The profile has a tail $\phi_b\simeq\sqrt{8/\lambda}\,R/\rho^2$ for $\rho\gg R$, so the weight of \refeq{Qdef} is not localized: $\rho^3\phi_b'\to-2\sqrt{8/\lambda}\,R$, a constant.

With $V''(\phi_b)=-3\lambda\phi_b^2=-24R^2/(\rho^2+R^2)^2$ and $u\equiv\eta-\ln R$,
\be\label{eq:FubK}\mathcal{K}&=&-\frac{d^2}{du^2}-6\,{\rm sech}^2u,\ee
a P\"oschl--Teller operator with $\nu=2$. This potential has two bound states and a continuum,
\be\label{eq:Fubspec}\kappa_0=-4,\;\;\hat F_0=\frac{\sqrt3}{2}\,{\rm sech}^2u;\;\;\;\;\;\;\kappa_1=-1,\;\;\hat F_1=\sqrt{\frac32}\,{\rm sech}\,u\,\tanh u;\;\;\;\;\;\;\kappa=k^2\geq0.\ee
$\hat F_0$ is \refeq{F0}, $r\phi_b'\propto{\rm sech}^2u$. The second bound state is the dilatation mode, $\partial_R\phi_b\propto(\rho^2-R^2)/(\rho^2+R^2)^2=F_1/r$. At $J=0$ its eigenvalue $\kappa_1+1$ vanishes, as scale invariance requires, while $F_0$ gives $\kappa_0+1=-3<0$, so that $\calO_{J=0}$ has a negative eigenvalue. At $J\geq2$ all eigenvalues $\kappa_n+(J+1)^2$ are positive. The ingredients of \refeq{Qspec} are elementary,
\be\label{eq:Fubdata}N_0=\frac{8}{3\lambda R^2},\;\;\;\;\;\Sigma_3=\frac{16}{3\lambda},\;\;\;\;\;\int d\eta\,\rho^2\hat F_0^2=2R^2,\;\;\;\;\;\int d\eta\,\rho^2\hat F_0\hat F_1=\frac{3\pi R^2}{2\sqrt2},\ee
the last from $\int du\,(e^{2u}-1)\,{\rm sech}^3u\,\tanh u=2\int du\,\sinh^2u/\cosh^3u=\pi$. (The identity \refeq{virial} gives $S_b=\pi^2\Sigma_3/2=8\pi^2/3\lambda$, as it should.) The wall-displacement part, \refeq{Qw}, is
\be\label{eq:Fubw}\calQ^2_{\rm wall}=\frac{3\pi}{32}\,\frac{\Sigma_3^2}{N_0}=\frac{\pi R^2}{\lambda}=\frac{3}{8\pi}\,S_bR^2.\ee
The $n=1$ term of \refeq{Qspec} has denominators $\kappa_1+(J+1)^2=J(J+2)$, and 
we find
\be\label{eq:Fub1}\calQ^2_{n=1}=N_0\left(\frac{3\pi R^2}{2\sqrt2}\right)^2\frac{2}{3\pi}=\frac{2\pi R^2}{\lambda}=2\,\calQ^2_{\rm wall}.\ee
Thus for this wide-wall potential the wall-displacement mode alone gives only part of the result: the first mode beyond the wall displacement already contributes twice as much. 

In fact, the leading parametric contribution in this model comes from the continuum modes. For $\rho,\rho'\gg R$ the weight in \refeq{Qdef} is the constant $-2\sqrt{8/\lambda}\,R$ and $G_J$ is the free massless Green's function $\rho_<^J/[2(J+1)\rho_>^{J+2}]$, so the double integral of \refeq{Qdef} over $R\ll\rho,\rho'\ll\rho_{\rm max}$ is $\ln(\rho_{\rm max}/R)/(J+1)^2$ for every $J$. This makes the quadrupole variance logarithmically divergent for $m\to0$. A false-vacuum mass cuts the integrals off at $\rho_{\rm max}\sim1/m$, and with
%
\be\label{eq:sumIR}\sum_{J=2,4,\dots}\frac{c_J}{(J+1)^2}&=&\frac{2}{\pi}\sum_{J=2,4,\dots}\left[\frac{1}{(J-1)(J+3)}-\frac{1}{(J-1)(J+3)(J+1)^2}\right]
\,=\,\frac{\pi}{16}\ee
the continuum gives
\be\label{eq:FubIR}\calQ^2_{\rm cont}=\frac{32R^2}{\lambda}\,\frac{\pi}{16}\,\ln\frac{1}{mR}+O\!\left(\frac{R^2}{\lambda}\right)=2\,\calQ^2_{\rm wall}\,\ln\frac{1}{mR}+\dots\,.\ee
The total result is therefore dominated by the continuum modes,
\be\label{eq:Fubtot}\calQ^2&=&\frac{2\pi R^2}{\lambda}\left[\ln\frac{1}{mR}+O(1)\right]\;\;\;\;\;\;(mR\ll1).\ee

\subsection{Light false vacuum with a sextic potential}\label{ssec:lfvnum}
To verify the coefficient of the logarithm in \refeq{Fubtot}, we evaluate \refeq{Qdef} for yet another family of potentials,
\be\label{eq:Ulfv}V(\phi)=\frac{M^4}{\lambda}\left[\frac{\delta^2}{2}x^2-\frac{1}{4}x^4+\frac{1}{6}x^6\right],\;\;\;\;\;x=\frac{\sqrt\lambda\,\phi}{M},\;\;\;\;\;\delta=\frac{m}{M},\ee
i.e., $V=\frac{1}{2}m^2\phi^2-\frac{\lambda}{4}\phi^4+\frac{\lambda^2}{6M^2}\phi^6$. For $\delta\to0$ the bounce is the Fubini profile with a radius fixed by the competition between the sextic term, which contributes $\propto1/(\lambda M^2R^2)$ to the action, and the mass term, which contributes $\propto m^2R^2\ln(1/mR)/\lambda$. This gives $MR\propto\delta^{-1/2}$ up to logarithms, so $mR\propto\delta^{1/2}\to0$ while $\phi_b(0)\propto1/R\to0$ keeps the sextic term a small perturbation at the core. The radial problems are solved numerically in the variable $\eta$. 
$R$ is defined by $x_b(R)=x_b(0)/2$, which would correspond to $R$ for the Fubini profile.

\reftab{lfv} gives the results for $\delta=10^{-2}$, $10^{-4}$ and $10^{-6}$, for which $mR=0.080$, $0.0060$ and $0.0005$. We see that, first, the wall-displacement part approaches its Fubini value \refeq{Fubw} at small $\delta$. Second, the dilatation mode contribution approaches \refeq{Fub1}. Third, we can study the logarithmic piece: between $\delta=10^{-2}$ and $10^{-6}$, $\ln(1/mR)$ increases by $5.02$ and $\calQ^2/\calQ^2_{\rm wall}$ increases by $10.01$, consistent with \refeq{FubIR}. Equivalently, the difference $\calQ^2/\calQ^2_{\rm wall}-2\ln(1/mR)$ stays constant, at $-1.20$, $-1.24$, $-1.22$, while $\calQ^2/\calQ^2_{\rm wall}$ itself grows from $3.8$ to $13.9$. Note that the constant residual piece (here $\sim-1.2$) is not universal, as it depends on how the potential cuts off the tail at $\rho\sim1/m$.

\begin{table}[t]
\centering
\sisetup{table-alignment-mode=none,group-digits=none}
\begin{tabular}{lSSSSS}
$\delta$ & {$mR$} & {$\dfrac{\lambda\calQ^2_{\rm wall}}{\pi R^2}$} & {$\dfrac{\calQ^2_{n=1}}{\calQ^2_{\rm wall}}$} & {$\dfrac{\calQ^2}{\calQ^2_{\rm wall}}$} & {$\dfrac{\calQ^2}{\calQ^2_{\rm wall}}-2\ln\dfrac{1}{mR}$}\\[4pt]
\hline
$10^{-2}$ & 0.0803 & 1.0113 & 1.077 & 3.85 & -1.20\\
$10^{-4}$ & 0.0060 & 0.9998 & 1.894 & 9.00 & -1.24\\
$10^{-6}$ & 0.0005 & 0.9999 & 1.990 & 13.86 & -1.22

\end{tabular}
\caption{The sextic family \refeq{Ulfv} at three values of $\delta=m/M$. The wall-displacement part in units of its Fubini value \refeq{Fubw}; the $n=1$ term of \refeq{Qspec}, summed over $J$, relative to it (Fubini value $2$, \refeq{Fub1}); the full result relative to it; and the same with the logarithm of \refeq{Fubtot} subtracted.}
\label{tab:lfv}
\end{table}

\section{Discussion and summary}\label{sec:disc}
The shape of a nucleating quantum phase transition bubble is drawn from a Gaussian distribution. 
For a thin-wall bubble, the quadrupole moment variance is of order $\hbar\sigma R^5$, and corresponds to a relative wall radius deformation 
\be\sqrt{\VEV{\delta R^2}}/R&\approx&(\hbar/S_b)^{1/2},\ee 
where $S_b$ is the bounce action. The right-hand side should be understood to have a model-dependent coefficient of order unity: in Sec.~\ref{sec:model} we survey model families that vary this coefficient between 1.2 and 2.6 or so. 
The scaling with $\hbar/S_b$ can be read off the thin wall. The wall is a membrane of tension $\sigma$, and the zero-point variance of its $L=2$ displacement is $\VEV{|\delta R_{2M}|^2}=3\pi\hbar/(32\sigma R)$. The tension and radius also set the bounce action, $S_b=\pi^2\sigma R^3/2$, so 
\be\label{eq:dRS}\frac{\VEV{|\delta R_{2M}|^2}}{R^2}&=&\frac{3\pi^3}{64}\,\frac{\hbar}{S_b}\;=\;(1.2)^2\,\frac{\hbar}{S_b}.\ee

For a cosmological phase transition the bounce action is related to the nucleation condition. With a nucleation rate per unit volume $\Gamma/V_3\sim\Lambda^4e^{-S_b/\hbar}$, $\Lambda$ being the energy scale of the sector undergoing the transition, bubbles appear and eventually percolate when $\Gamma/V_3\sim H^4$, with $H$ the Hubble parameter. Thus a successful transition satisfies $S_b/\hbar\simeq4\ln(\Lambda/H)$. With $H=1.66\sqrt{g_*}\,T^2/M_{\rm Pl}$ and $\Lambda\sim T$, this gives $S_b/\hbar\simeq120$--$150$ for a transition occurring at a temperature $T$ between $100$~TeV and $100$~GeV. Hence the $L=2$ deformation of a nucleating bubble wall is $\delta R/R\simeq0.1$ per mode. \reffig{popcorn} shows what this looks like, including also higher multipoles. 

For a thin wall the quadrupole is dominated by the $L=2$ modulation of the wall treated as a membrane with no internal structure. For a thick wall, internal modes in the wall also contribute. For a light false vacuum, at $mR\ll1$ where $m$ is the mass of field fluctuations at the false vacuum, the quadrupole is dominated by the field fluctuations on the $\rho^{-2}$ tail of the bubble, leading to a parametric correction of $2\ln(1/mR)$. In this limit the quadrupole is dominated by massless fluctuations traversing the bubble.

It could be interesting to extend the gravitational-wave computation of~\cite{Blum:2024hcs} to the time span in the history of the bubble near its nucleation. 
We do not pursue this here, but our calculations demonstrate how to use the state of the field on the nucleation slice, \refeq{PsiT}. We can also extend our analysis to incorporate time evolution: this can be achieved by the analytic continuation of the Euclidean two-point function in both time arguments, $\tau\to it$, $\tau'\to it'$, with the bounce continued to the expanding bubble $\phi_b(\sqrt{\rho^2-t^2})$. Equivalently, the Gaussian state \refeq{PsiT} can be evolved with the linearized field equation on the expanding bubble, with mode functions whose initial data at $t=0$ are fixed by $M$. For the wall-displacement mode of an advancing thin wall, this procedure matches the mode-function construction of Ref.~\cite{Blum:2024hcs}. 

\acknowledgments
We thank Mehrdad Mirbabyi for useful discussions at an early stage of this work. IF and GLP are supported by the ERC (NOTIMEFORCOSMO, 101126304), by Scuola Normale, and by INFN (IS GSS-Pi). GLP is also supported by the Italian Ministry of Universities and Research (MUR) under contract 20223ANFHR (PRIN2022). 

\begin{appendix}
\section{Hyperspherical reduction}\label{app:hsh}
The hyperspherical harmonics on $S^3$ satisfy the addition theorem
\be\label{eq:addition}\sum_{L=0}^{J}\sum_{M=-L}^{L}\mathcal{Y}_{JLM}(\hat x)\,\mathcal{Y}^*_{JLM}(\hat y)&=&\frac{J+1}{2\pi^2}\,U_J(\hat x\cdot\hat y),\ee
which is consistent with the degeneracy $(J+1)^2=U_J(1)(J+1)$ and the volume $2\pi^2$ of $S^3$, and gives the completeness relation $\delta_{S^3}(\hat x,\hat y)=\sum_J\frac{J+1}{2\pi^2}U_J(\hat x\cdot\hat y)$. With $\delta^4(x-y)=r^{-3}\delta(r-r')\,\delta_{S^3}$ and \refeq{GJdef}, this gives \refeq{GEdec}. As a check, for $V''=0$ the radial Green's function is $G_J=r_<^J/[2(J+1)r_>^{J+2}]$ and the generating function $\sum_Jt^JU_J(c)=(1-2tc+t^2)^{-1}$ reproduces $G_E=1/(4\pi^2|x-y|^2)$.

With the harmonics \refeq{YJLM} on the slice, $\mathcal{Y}_{JLM}=N_{JL}Y_{LM}(\hat n)$ with $N_{JL}\equiv N_{JL}(\pi/2)$. Comparing \refeq{addition} at $\beta=\beta'=\pi/2$ with the Legendre expansion \refeq{UP} and $\sum_MY_{LM}Y^*_{LM}=\frac{2L+1}{4\pi}P_L$ gives $u_{JL}=\frac{\pi(2L+1)}{2(J+1)}N_{JL}^2$. For $L=2$ and $J=2k+2$, $C^3_{2k}(0)=(-1)^k(k+1)(k+2)/2$, and
\be\label{eq:cJapp}c_J=\frac{2(J+1)}{5\pi}u_{J2}=\frac{128(J+1)}{\pi}\frac{(J-2)!}{(J+3)!}\frac{J^2(J+2)^2}{64}=\frac{2}{\pi}\frac{J(J+2)}{(J-1)(J+3)},\ee
which is \refeq{cJ}. 

\section{Making popcorn}\label{app:popcorn}
\reffig{popcorn} is a draw from the Gaussian ensemble of wall displacements in the thin-wall limit. As we explain below, it is  ``observed'' at a finite resolution, namely, calculated for a smeared field. 

Define the displacement of the wall in each harmonic by the projection of the slice field on the wall mode,
\be\label{eq:dRLM}\delta R_{LM}&\equiv&-\frac{1}{N_0}\int_0^\infty d\rho\,\rho\,\phi_b'(\rho)\,\dphi_{LM}(\rho),\ee
with $\dphi_{LM}$ of \refeq{dphiLM} and $N_0$ of \refeq{Qspec}.
With $d\rho=\rho\,d\eta$, \refeq{dRLM} is $-\frac{1}{N_0}\int d\eta\,F_0\,\rho\,\dphi_{LM}$, and by \refeq{gL} and \refeq{GJspec} the covariance of $\rho\,\dphi_{LM}$ is $\hbar\sum_Jc^{(L)}_J\sum_n\hat F_n(\eta)\hat F_n(\eta')/[\kappa_n+(J+1)^2]$. The radial integrals project on the ground state of $\mathcal{K}$ alone, $\int d\eta\,F_0\hat F_n=\sqrt{N_0}\,\delta_{n0}$, so
\be\label{eq:sL}\VEV{|\delta R_{LM}|^2}&=&\frac{\hbar}{N_0}\,s_L,\;\;\;\;\;\;s_L=\sum_{J=L,L+2,\dots}\frac{c^{(L)}_J}{(J-1)(J+3)},\ee
with $c^{(L)}_J$ of \refeq{cLJ}. At $L=2$ the sum is the one evaluated below \refeq{Qw}, $s_2=3\pi/32$, and \refeq{sL} gives \refeq{dRw}. In the thin-wall limit
\be\label{eq:sLthin}\frac{\VEV{|\delta R_{LM}|^2}}{R^2}&=&\frac{\pi^2}{2}\,\frac{\hbar}{S_b}\,s_L.\ee
For $L=2$, $c^{(2)}_J=c_J$ has the closed form \refeq{cJ}. For general $L$ we evaluate $u_{JL}$ of \refeq{UP} by Gauss--Legendre quadrature, 
sum $J$ up to $2\times10^4$, and add the tail of the sum analytically, using $c^{(L)}_J\to2/\pi$ at large $J$.
We depict graphically $L\geq2$ only, as we are not interested in radial fluctuations or fluctuations that amount to a displacement of the bubble center. Before we proceed to make a plot, however, we must properly define an observable smeared field.

The displacement at a point is $\delta R(\hat n)=\sum_{LM}\delta R_{LM}Y_{LM}(\hat n)$, and by \refeq{sL} its variance is
\be\label{eq:ptvar}\VEV{\delta R(\hat n)^2}&=&\frac{\hbar}{N_0}\sum_{L\geq2}\frac{2L+1}{4\pi}\,s_L.\ee
This variance diverges linearly in the maximal $L$, reflecting the fact that the variance of a free field at a point diverges with the momentum cutoff. A picture of the bubble, like a picture of any field, must be defined together with the resolution at which the field is observed. We define the observed field by smoothing with a Gaussian of width $\ell$,
\be\label{eq:smooth}\phi_\ell(\bx)&=&W_{\ell}*\phi(\bx)\,=\,\int d^3x'\,W_\ell(\bx-\bx')\,\phi(\bx'),\;\;\;\;\;\;W_\ell(\bx)\,=\,\frac{e^{-\bx^2/2\ell^2}}{(2\pi\ell^2)^{3/2}}.\ee
For the bounce itself, $\phi_{b,\ell}=W_\ell*\phi_b$. The observed wall is the wall of $\phi_{b,\ell}$ and its displacement is 
\be\label{eq:dRLMell}\delta R_{\ell,LM}&\equiv&-\frac{1}{N_{0,\ell}}\int_0^\infty d\rho\,\rho\,\phi_{b,\ell}'(\rho)\,(\dphi_\ell)_{LM}(\rho),\;\;\;\;\;\;N_{0,\ell}\,=\,\int_0^\infty d\rho\,\rho\,\phi_{b,\ell}'(\rho)^2.\ee
%
Write $\bx=\rho\hat n$ and $\bx'=\rho'\hat n'$, so that $W_\ell(\bx-\bx')=(2\pi\ell^2)^{-3/2}e^{-(\rho^2+\rho'^2)/2\ell^2}e^{z\,\hat n\cdot\hat n'}$ with $z=\rho\rho'/\ell^2$, and expand 
\be\label{eq:iL}e^{z\,\hat n\cdot\hat n'}&=&\sum_{L=0}^\infty(2L+1)\,i_L(z)\,P_L(\hat n\cdot\hat n'),\no\\
i_L(z)&=&\sqrt{\frac{\pi}{2z}}\,I_{L+\frac12}(z)\,=\,\frac{e^z}{2z}\,e^{-L(L+1)/2z}\left[1+O\!\left(\frac{L^2}{z^2}\right)\right],\ee
with $i_L(z)$ the modified spherical Bessel function. With $\dphi(\bx')=\sum_{LM}\dphi_{LM}(\rho')Y_{LM}(\hat n')$ in \refeq{smooth}, the angular integral is $\int d\Omega'\,P_L(\hat n\cdot\hat n')\,Y_{LM}(\hat n')=\frac{4\pi}{2L+1}Y_{LM}(\hat n)$, and the moments of the smoothed field are
\be\label{eq:WY}(\dphi_\ell)_{LM}(\rho)&=&\int_0^\infty d\rho'\,\frac{\rho'}{\rho}\,\frac{e^{-(\rho-\rho')^2/2\ell^2}}{\sqrt{2\pi\ell^2}}\,e^{-L(L+1)\ell^2/2\rho\rho'}\,\dphi_{LM}(\rho'),\ee
up to the correction in \refeq{iL}. 

The smoothed bounce is \refeq{WY} at $L=0$ with $\phi_b$ in place of $\dphi_{00}$. Differentiating with respect to $\rho$ and integrating by parts gives
\be\label{eq:den}\phi_{b,\ell}'(\rho)&=&\int_0^\infty d\rho'\,\frac{\rho'}{\rho}\,\frac{e^{-(\rho-\rho')^2/2\ell^2}}{\sqrt{2\pi\ell^2}}\,\phi_b'(\rho')\left[1+O\!\left(\frac{\ell}{R}\right)\right].\ee
Now insert the slice field of the ensemble, $\dphi_{LM}(\rho')=-\phi_b'(\rho')\,\delta R_{LM}$, into \refeq{WY}. The Gaussian and $\phi_b'$ confine $\rho$ and $\rho'$ to within a few $\ell$ of $R$, where $\rho\rho'=R^2[1+O(\ell/R)]$ and $z\simeq R^2/\ell^2$. The angular factor is therefore
\be\label{eq:wL}w_L&=&e^{-L(L+1)\ell^2/2R^2},\ee
up to a relative correction $O(\ell/R)$ for every $L$ at which $w_L$ is not negligible, $L\lesssim R/\ell$. 
What remains of \refeq{WY} is \refeq{den},
\be\label{eq:smwall}(\dphi_\ell)_{LM}(\rho)&=&-w_L\,\delta R_{LM}\,\phi_{b,\ell}'(\rho)\left[1+O\!\left(\frac{\ell}{R}\right)\right].\ee
The smoothed field of a displaced wall is the smoothed wall, displaced by $w_L\delta R_{LM}$. Inserting \refeq{smwall} in \refeq{dRLMell},
\be\label{eq:dRellres}\delta R_{\ell,LM}&=&w_L\,\delta R_{LM}.\ee
The smoothing acts on the position of a thin wall through its angular part only, the heat kernel on the sphere of angular width $\ell/R$. Its radial part drops out of \refeq{dRLMell}, the projection on the wall profile being already a smoothing over the width of the wall.

What we draw is the observed wall,
\be\label{eq:popsurf}\rho(\hat n)&=&R\Big[1+\sum_{L\geq2}\sum_{M}w_L\,a_{LM}\,Y_{LM}(\hat n)\Big],\ee
with real orthonormal $Y_{LM}$ and $a_{LM}$ independent Gaussian amplitudes of variance \refeq{sLthin}, at $S_b/\hbar=100$. The sum over $L$ is carried out until $w_L^2<10^{-11}$, which for $\ell=R/10$ is $L=50$. The variance of the observed displacement at a point is \refeq{ptvar} with $w_L^2$ inside the sum,
\be\label{eq:poprms}\frac{\VEV{\delta R_\ell(\hat n)^2}}{R^2}&=&\frac{\pi^2}{2}\,\frac{\hbar}{S_b}\sum_{L\geq2}\frac{2L+1}{4\pi}\,w_L^2\,s_L\;\simeq\;\frac{\pi^2}{16\sqrt\pi}\,\frac{\hbar}{S_b}\,\frac{R}{\ell},\ee
the last form for $\ell\ll R$, from $s_L\to1/2L$. At $S_b/\hbar=100$ and $\ell=R/10$ the sum gives $\delta R_\ell/R=0.17$. \refeq{poprms} is the bubble as seen at resolution $\ell$, in the sense in which $\VEV{\dphi_\ell^2}\propto\hbar/\ell^2$ is a free field as seen at resolution $\ell$. The content that does not depend on the resolution is the spectrum \refeqs{sL}{sLthin}, the variance of every multipole, and the growth of \refeq{poprms} with $R/\ell$ is the statement that the wall is rough on every scale between $R$ and $\ell_w$. The shape of the kernel enters only the coefficient: a sharp cutoff at $L=R/\ell$ gives $\pi/8$ in place of $\pi^2/16\sqrt\pi$. For a wall of finite width $\ell_w$ the membrane description of the displacement holds for $L\lesssim R/\ell_w$, and the figure assumes $\ell\gg\ell_w$.
Note that the quadrupole moment of the bubble needs no smoothing, as it corresponds to a single angular component.

Any $L=2$ deformation is $\delta R/R=\hat n\cdot A\cdot\hat n$ with $A$ a traceless symmetric matrix, that is, to first order, a triaxial ellipsoid with semi-axes $R(1+e_i)$, $\sum_ie_i=0$. The sample of \reffig{popcorn} is drawn in the principal frame of its $L=2$ part, with $e_i=(0.13,\,0.01,\,-0.14)$. Its five $L=2$ amplitudes are the first random draw with $\sum_Ma_{2M}^2$ within $10\%$ of its mean, so that the ellipsoid is typical rather than lucky, and the amplitudes at $L\geq3$ are the following draws. Don't forget to add butter and salt.
\end{appendix}

\bibliography{ref}
\bibliographystyle{utphys}

\end{document}